\documentclass[pre,superscriptaddress,showpacks]{revtex4-1}

\usepackage{amsfonts}
\usepackage{amssymb}
\usepackage{graphicx}
\usepackage{color}
\usepackage{amsthm}
\usepackage{amsmath}

\newcommand{\ii}{\imath}

\newcommand{\half}{\frac{1}{2}}
\newcommand{\Tr}{\textrm{Tr}}

\newcommand{\f}[1]{\textrm{#1}}

\newcommand{\ev}[1]{\overline{#1}}

\begin{document}

\title{Spin decoherence due to Fluctuating Fields}

\author{Piotr Sza\'{n}kowski}
\affiliation{Institute of Theoretical Physics, University of
Warsaw, ul.  Ho\.{z}a 69, PL--00--681 Warszawa, Poland}
\affiliation{Department of Chemistry, Department of Physics and
Department of Electro-Optics, and the Ilse Katz Center for Nano-Science,
Ben-Gurion University, Beer-Sheva 84105, Israel}
\author{M. Trippenbach}
\affiliation{Institute of Theoretical Physics, University of
Warsaw, ul.  Ho\.{z}a 69, PL--00--681 Warszawa, Poland}
\author{Y. B. Band}
\affiliation{Department of Chemistry, Department of Physics and
Department of Electro-Optics, and the Ilse Katz Center for Nano-Science,
Ben-Gurion University, Beer-Sheva 84105, Israel}


\begin{abstract}
The dynamics of a spin in the presence of a deterministic and a
fluctuating magnetic field is solved for analytically to obtain the
averaged value of the spin as a function of time for various kinds of
fluctuations (noise).  Specifically, analytic results are obtained for
the time dependence of the expectation value of the spin, averaged
over fluctuations, for Gaussian white noise, Guassian colored noise,
as well as non-Gaussian telegraph noise.  Fluctuations cause the decay
of the average spin vector (decoherence).  For noise with finite
temporal correlation time, a deterministic component of the field can
suppress decoherence of the spin component along the field.  Hence,
decoherence can be manipulated by controlling the deterministic
magnetic field.  A simple universal physical picture emerges which
explains the mechanism for the suppression of decay.
\end{abstract}


\maketitle

\section{Introduction}  \label{Sec:Intro}

Quantum systems are almost invariably coupled to an environment that
degrades their coherence.  Decoherence of quantum systems is a
fundamental problem, with implications across all branches of physics.
The temporal evolution of a spin system in the presence of coupling to
an environment is perhaps the simplest phenomenon involving
decoherence.  For example, a particle of spin 1/2, which can be mapped
onto any two-level system (TLS) \cite{Feynman_63}, and {\it vice
versa}, has served as an important paradigm \cite{Itakura_03,
Altshuler_09, Cheng_08, Aharony_10}.  Specific examples include the
decoherence of a qubit in a quantum-computer or quantum-information
processor \cite{QI}, spin dynamics in a magnetometer \cite{Budker_07},
accuracy and stability limitations of atomic clocks due to
interactions with an environment \cite{clocks}, decoherence of
two-level quantum dot systems \cite{QDots, TaylorQDots}, nitrogen
vacancy centers in diamond \cite{NV, NV_fluc}.

Under certain assumptions, which we discuss in the next section, the
evolution of a spin in the presence of an environment can be
represented by evolving the spin in an effective magnetic field, $\vec
B^{(\f{eff})} = \vec B_0 + \vec B_E(t)$.  Here $\vec B_0$ is a
deterministic magnetic field (which can be time-dependent) that can be
used to exert control over the spin.  The magnetic field $\vec B_E(t)$
models the influence of an environment on the system, and is
represented by a vector stochastic process $\vec b(t)$.  Averaging
over fluctuations corresponds to tracing out the environmental degrees
of freedom.  This yields a reduced, non-unitary dynamics wherein the
averaged spin decoheres in time.

The physical properties of the environment determine the statistical
properties of $\vec B_E(t)$, which in turn determine the type of
stochastic process $\vec b(t)$.  Here we consider several types of
stochastic fluctuations.  A prototype model for fluctuations is
Gaussian white noise \cite{vanKampenBook, Kloeden}, a random process
having vanishing correlation time.  As a consequence, the system
driven by it evolves without {\it memory}, since it has no way to
probe the past.  A natural generalization is a Gaussian colored noise
\cite{Valsakuma_85, vanKampen_89}, wherein the random process has a
finite correlation time, hence, the system evolves with memory.  The
ubiquitousness of this type of noise can be attributed to central
limit theorem which states that the superposition of a very large
collection of independent sources of fluctuation tends to Gaussian
noise.  Moreover, in some spin systems there are more than one source
of noise which differ in their intensity and correlation time scale.
For example, in the nitrogen-vacancy diamond system
\cite{Bar-Gill_12}, NV centers experience slowly fluctuating noise
from $^{13}$C impurity nuclear magnetic moments, and extremely fast
fluctuations from nearby electronic nitrogen spins.  The former can be
modeled by colored noise, which is characterized by a very long
correlation time, while the latter can be treated as a white noise,
since it sets the fastest time scale in the system.  A particular case
of colored noise that is {\em not} Gaussian, telegraph noise, has been
studied in connection with quantum dot qubit systems \cite{Aharony_10,
Altshuler_09}.  Reference \cite{Cheng_08} claims that telegraph noise
can be used to model the environment of systems where a spin can
interact with relatively few random fluctuators in its neighborhood,
and these fluctuators go back and forth between only two states.

The idea of using stochastic methods for modeling an environment or
other complex interactions within the spin system is not new.  For
example, the problem of the line shape in paramagnetic resonance is
discussed in Ref.~\cite{Anderson_53}.  It was assumed that the
electrons in an atom are affected by a perturbation which varies
randomly in time at a rate determined by exchange interactions.  As a
result, the fluctuations around the stationary state are induced and
their effect on the absorption spectra can be calculated.  Another
example is presented in Ref.~\cite{Anderson_62}, where the authors
consider an ensemble of spins interacting with an environment as well
as each other.  The problem of calculating the effects of these
complex interactions on the observed line shape is circumvented by
introducing stochastic fluctuations of the precession frequency of the
observed spins.

The goal of this work is to develop methods to describe and analyze
the process of relaxation of a single spin induced by contact with an
environment.  An important result of our analysis is that, when the
spin evolves with memory, it can be manipulated by applying a
sufficiently strong deterministic magnetic field $\vec B_0$ to
significantly suppress decoherence.  We hope that the level of
generality and overall simplicity of our approach allows for a
straightforward application to appropriate experimental systems.

The outline of the this paper is as follows.  Section
\ref{Sec:Spin_dyamics} develops the equations of motion for a spin in
the presence of a deterministic magnetic field and a stochastically
fluctuating magnetic field.  Section~\ref{Sec:GCN} describes evolution
of the system with Gaussian colored noise fluctuations.  In that
section we present a simple, intuitive view of the nature of the decay
of the averaged spin due to field fluctuations.  We also show that
this decay can be suppressed when the deterministic magnetic field is
strong enough, and the correlation time is long enough (i.e., when the
system has memory).  Section~\ref{Sec:white-noise} develops the white
noise limit of colored noise.  Section~\ref{Sec:Memory} explains the
effects of memory on decoherence, and Sec.~\ref{Sec:Telegraph}
considers telegraph noise, which is a specific from of non-Gaussian
noise.  Finally, a summary and a conclusion are presented in
Sec.~\ref{Sec:Summary}.

\section{Spin dynamics in a fluctuating magnetic field}
\label{Sec:Spin_dyamics}

We consider fluctuations in an otherwise deterministic system that
results from the application of a effective random field generated by
an environment with which it interacts.  The back-coupling of the spin
to the source of the noise is not taken into account (Van Kampen
\cite{vanKampenBook} calls this {\it external noise}).  Hence, the
stochastic properties of the noise result only from the environment
which is unaffected by the system.  We assume that these properties
can be measured or otherwise deduced.  For each particular system, one
should carefully check whether the back-action can indeed be
neglected.  Examples of systems for which back-action was neglected
include: nitrogen-vacancy centers in diamond affected by magnetic
moments of impurity spins \cite{Lange,NV_fluc}, electron spin in a
quantum dot affected by nuclear magnetic moments that fluctuate due to
crystal lattice vibrations \cite{TaylorQDots,Dobrovitski}, and
magnetic noise in atom chips caused by fluctuations of electron
currents in wires that make up the atom chip \cite{Folman_05}.

Consider a particle of spin ${\vec S}$ with magnetic moment $\vec\mu
=\mu \vec S$ in the presence of a time-dependent field $\vec B(t)$.
The dynamics is determined using a Zeeman-type Hamiltonian,
\begin{equation}  \label{eq:H_Zeeman}
    H = -\vec{\mu} \cdot \vec B(t) =  \vec\Omega(t)
    \cdot \vec S ,
\end{equation}
where $\vec S$ is the vector of spin operators satisfying commutation
relations $[ S_i , S_j ] =i \hbar \epsilon_{ijk}S_k$, $\vec\Omega(t) =
-\mu \vec B(t)/\hbar$ is the Rabi frequency vector, $\vec B(t)$ is a
total magnetic field felt by the particle, $\mu = g \mu_0$ is the
magnetic moment of the particle, $g$ is the $g$-factor and $\mu_0$ is
the Bohr (or nuclear) magneton.  The magnetic field ${\vec B}(t)$ is
the sum of a deterministic field, $\vec B_0$, whose direction defines
the $z$-axis, and a fluctuating field, $\vec b(t)$.  Hence, the Rabi
frequency is given by
\begin{equation}
    \vec{\Omega}(t) = \left( \! \begin{array}{c} 0 \\0 \\ \Omega_0(t) \\
    \end{array} \! \right) + \left( \! \begin{array}{c}
    \omega_x(t)\\\omega_y(t)\\\omega_z(t)\\
    \end{array} \! \right) ,
\end{equation}
where $\vec\Omega_0 = -\mu \vec B_0/\hbar$ and $\vec\omega(t) = -\mu
\vec b(t)/\hbar$.  For simplicity, we have taken the deterministic
field to have a component only along the $z$-axis.  If the
deterministic field varies slowly in time (adiabatically), our
conclusions below remain valid, but for deterministic fields that vary
more quickly, the simple version of the deterministic part of the
evolution matrix that satisfies Eq.~(\ref{eq:U}) requires revision, as
detailed below.

In the Heisenberg representation, the evolution of the spin
operators is given by the equations,
\begin{equation}  \label{eq:Heisenberg}
    \frac{d}{d t} S_i^{(H)}(t)=\frac{\ii}{\hbar} \left[ H(t),
    S_i^{(H)}(t)\right] ,
\end{equation}
where the superscript $(H)$ denotes the Heisenberg picture.  In our
case, the Hamiltonian contains a stochastic term, $\vec\omega(t)$:
\begin{equation} \label{eq:Hamiltonian}
    H(t) = \Omega_0(t) S_z + \vec\omega(t) \cdot \vec S .
\end{equation}
The commutator on the right hand side of the Heisenberg equation of
motion, Eq.~(\ref{eq:Heisenberg}), is easy to evaluate, and leads to
the vector form of the equation,
\begin{equation}  \label{eq:Bloch}
    \frac{d}{d t}\vec{S}^{(H)}(t) = \left\{\Omega_0(t) {\vec e}_z +
    \vec\omega(t)\right\} \times \vec{S}^{(H)}(t) ,
\end{equation}
and the initial condition is simply
\begin{equation}  \label{eq:initial}
    \vec{S}^{(H)}(0) = \left(\begin{array}{c} S_x\\
    S_y \\ S_z \\ \end{array}\right) .
\end{equation}
The solution to Eq.~(\ref{eq:Bloch}) with the initial condition
(\ref{eq:initial}) can be written as a 3$\times$3 evolution matrix,
$\hat U(t,0)$, acting on the initial vector, $\vec{S}^{(H)}(0)$.  It
is convenient to write the equations, $S_i^{(H)}(t) = \sum_{j=1}^3
U_{ij}(t,0) S_j^{(H)}(0)$, in vector notation as $\vec{S}^{(H)}(t) =
\hat U(t,0) \vec{S}^{(H)}(0)$, where the hat ${\hat \,}$ indicates a
3$\times$3 matrix.  For convenience, let us take the deterministic
field $\Omega_0$ to be constant in time.  Substituting into the
equations of motion yields,
\begin{equation}  \label{eq:U}
    \frac{d}{d t}\hat U(t,0) = \left( \Omega_0 \, \hat\varepsilon^z +
    \sum_{k=1}^3\omega_k(t) {\hat\varepsilon^k}\,\right) \hat
    U(t,0).
\end{equation}
Here the matrix elements of $\hat\varepsilon^ k$ are defined as
$\varepsilon^k_{ij} \equiv \epsilon_{ikj}$, where $\epsilon_{ikj}$ is
the Levi-Civita symbol.  The matrices $\hat \varepsilon^i$ satisfy
commutation relations similar to those of angular momentum operators,
\begin{equation}\label{eq:EpsCommRel}
    [ \hat \varepsilon^i,\hat\varepsilon^j ] = \epsilon_{ijk}
    \hat\varepsilon^k.
\end{equation}
Consequently, they are generators of rotations in a real,
three-dimensional vector space.  For example, $\hat\varepsilon^z$
generates rotations around $z$ axis,
\begin{equation}\label{eq:rotation_matrix}
    \hat R(\theta , z) \equiv \exp\{ \theta \hat\varepsilon^z \} =
    \left(\begin{array}{ccc} \cos\theta & -\sin\theta & 0 \\
    \sin\theta &\cos\theta & 0 \\
    0 & 0 & 1 \\
  \end{array}\right) .
\end{equation}
The evolution operator $\hat U$ is a rotation operator that causes
precession of the spin vector around the instantaneous rotation axis
defined by the total field $\vec{\Omega}(t) = \vec\Omega_0 +
\vec\omega(t)$.  The equation of motion for the spin does not depend
upon the details of the representation of the spin, hence the solution
to the dynamics for an arbitrary spin is also given by (\ref{eq:U}).
The only indication of the spin representation, i.e., the dimension
$2S+1$ of the representation, is in the initial condition.

Experimental measurements of the spin will of necessity correspond to
quantum expectation values and averages over the stochastic
fluctuations of the magnetic field.  Hence the quantity $\langle
\ev{{\vec S}(t)} \rangle$ corresponds to experimental measurements of
the spin.  Here, the symbol $\langle \, \, \rangle$ means quantum
average, and the symbol $\ev{ {\color{white} S} }$ means average over
stochastic variables.  Note that the average over fluctuations and the
average over the initial quantum state factorizes in the following
way: $\langle \ev{{\vec S}(t)} \rangle=\ev{\hat U(t,0)}\langle\vec
S(0)\rangle$.

For spin 1/2 (i.e., for a TLS), the average expectation value of the
spin ${\vec S} = \hbar {\vec \sigma}/2$ fully determines the state of
the system.  This is because the density matrix $\varrho$ can be
expressed using the Pauli matrices ${\vec \sigma}$ which, together
with identity operator $1$, form a basis in the space of 2$\times$2
hermitian matrices.  In the Schr\"{o}dinger representation,
\begin{equation}\label{eq:av_density_matrix}
    \overline{\varrho(t)} =\half\left( 1 + \Tr\left\{\ev{{\vec
    \sigma}^{(H)}(t)}\varrho(0)\right\} \cdot {\vec \sigma}\right)
    =\half\left( 1 + \overline{ \langle \vec \sigma(t) \rangle} \cdot
    \vec\sigma \right) .
\end{equation}
and the expectation values of the spin at time $t$ is given by vector
parameter ${\vec \lambda}(t) \equiv \langle \ev{{\vec \sigma}(t)}
\rangle$.  Thus, at any time $t$, there are three parameters,
$\lambda_i(t)$, $i = x, y, z$, which completely specify the density
matrix.  This does not mean that the dynamics of the density matrix
for any $(2S+1)$-level system can be represented in terms of the
dynamics of a particle with spin ${\bf S}$ coupled to a magnetic
field, since the number of parameters necessary to specify the density
matrix larger than the three components of the spin
\cite{No_indep_param}.

\section{Gaussian Colored Noise}  \label{Sec:GCN}

First, let us consider a process that is Gaussian, isotropic and
Markovian \cite{vanKampenBook}.  The field fluctuates in all three
dimensions with independent components (isotropy) and statistical
properties of each component is completely determined by the first two
moments (Gaussian), the average and the correlation function.  The
stationary condition means that the correlation function depends on
the time difference and the average is time independent.  According to
Doob's theorem \cite{Doob}, if the process is also Markovian, the
correlation function is exponential and the average is zero, i.e.,
\begin{eqnarray}
    \overline{\omega_i(t)}&=&0 , \\
    \overline{\omega_i(t)\omega_j(t')} &=& \kappa(t-t') \, \delta_{ij}
    = \omega_0^2 \, e^{-\frac{|t-t'|}{\tau_c}}\,\delta_{ij} .
\end{eqnarray}
Here the bar indicates average over the fluctuations, $\kappa(t-t')$
is the correlation function with correlation time $\tau_c$, the
average fluctuating Rabi frequency vanishes, $\overline{\vec \omega} =
0$, and the variance of the Rabi frequency, $\omega_0^2$, specifies
the strength of the fluctuations.

Several comments regarding Guassian colored noise are in order.
First, note that the Gaussian process can model a wide variety of
environments due to the central limit theorem; if the environment
consists of many individual and independent fluctuating elements, the
central limit theorem substantiates the approximation of using a
Gaussian process as the effective field affecting the spin.  Secondly,
if there is no back-action of the spin on the environment, and the
environment is in a stationary state (e.g., is in equilibrium) then
the process is stationary as well.  Finally, the isotropy assumption
regarding the random magnetic field produced by the environment is
appropriate if no particular special direction can be associated with
the environment.

To solve the equation of motion (\ref{eq:U}), first we eliminate the
so called \emph{drift term} due to the deterministic part of the
magnetic field, $ \Omega_0 \, \hat\varepsilon^z \hat U(t,0)$, from the
right hand side of the equation by transforming to the rotating frame.
Then Eq.~(\ref{eq:U}) takes the form
\begin{equation}  \label{eq:UPrime}
    \frac{d}{d t}\hat U'(t,0) = \left(\vec\omega(t) \cdot
    \vec{\hat\varepsilon}'(t)\right) \hat U'(t,0).
\end{equation}
In this equation we introduced transformed matrices, denoted by a
prime, $\vec{\hat\varepsilon}'(t)$ and $\hat U'(t,0)$, which are given
by
\begin{eqnarray}
(\varepsilon ')^k_{i j}(t) & \equiv & \sum_{m,n}R_{i m}(t\Omega_0,z) \,
\varepsilon^k_{m n} \, R_{n j}(-t\Omega_0,z) =
R_{k l}(t\Omega_0,z) \,\varepsilon^l_{i j} , \\
U_{i j}'(t_1,t_2)  &\equiv& \sum_{m,n}R_{i m}(t_1\Omega_0,z) \,
U_{m n}(t_1,t_2) \, R_{n j}(-t_2\Omega_0,z) ,
\end{eqnarray}
where $\hat R(t\Omega_0,z) = \exp\{ t\Omega_0 \hat \varepsilon^z\}$ is
a deterministic Euclidean rotation matrix that results in
counterclockwise rotation around the $z$-axis by the angle $\Omega_0
t$ [see Eq.~(\ref{eq:rotation_matrix})].  Note that the scalar product
$\vec\omega(t) \cdot \vec{\hat\varepsilon}'(t)$ in
Eq.~(\ref{eq:UPrime}) can be rewritten as
\begin{equation}
    \vec\omega(t)\cdot\vec{\hat\varepsilon}'(t) = \left(\hat
    R(-t\,\Omega_0,z) \, \vec\omega(t)\right)\cdot\vec{\hat\varepsilon}
    \equiv \vec\omega'(t) \cdot \vec{\hat\varepsilon} .
\end{equation}
The time-dependent rotation matrix $\hat R(t\,\Omega_0,z)$ ``mixes''
field fluctuations in the $x$ and $y$ component.  In the Lab frame,
$\omega_x$ and $\omega_y$ are independent, however, because of the
mixing, they are correlated in the rotating frame and the correlation
functions are modified in the following way:
\begin{eqnarray}
 \overline{\omega_{x}'(t_1)\omega_{y}'(t_2)} &=& -
 \overline{\omega_{y}'(t_1)\omega_{x}'(t_2)} = \kappa(t_1-t_2)
 \sin\left[\Omega_0(t_1-t_2)\right],\label{eq:kappa_XY}\\
 \overline{\omega_{x}'(t_1)\omega_{x}'(t_2)} &=&
 \overline{\omega_{y}'(t_1)\omega_{y}'(t_2)} =
 \kappa(t_1-t_2)\cos\left[\Omega_0(t_1-t_2)\right] , \label{eq:kappa_XX}\\
 \overline{\omega_{z}'(t_1)\omega_{i}'(t_2)} &=&
 \kappa(t_1-t_2) \, \delta_{3 i} .\label{eq:kappa_ZZ}
\end{eqnarray}
The stochastic field in the rotating frame, $\vec\omega'(t)$, remains
a Gaussian process, since it is a linear combination of Gaussian
processes.  Note that any time-independent rotation does not change
the correlation functions of the isotropic Gaussian vector process
$\vec\omega(t)$.

The formal solution to Eq.~(\ref{eq:UPrime}) is given in the terms of the
time-ordered exponential function,
\begin{equation}  \label{eq:Texp}
    \hat U'(t,0) = \mathcal T \exp\left\{ \sum_i \hat\varepsilon^i
    \int_0^t\omega_i '(\tau)d\tau \right\} ,
\end{equation}
defined by the power series, $\mathcal T \exp \int_0^t \hat
A(\tau)d\tau \equiv \hat 1 + \int_0^t d\tau_1\hat A(\tau_1) +
\int_0^td\tau_1 \int_0^{\tau_1} d\tau_2 \, \hat A(\tau_1)\hat
A(\tau_2)+\ldots$, where $\mathcal T$ is the time-ordering operator
\cite{vanKampen_Cumulant_Exp}.  The average value of $\hat U'$,
$\overline{\hat U'(t,0)}$, can be computed by noting that it has the
form of a {\it moment generating functional} \{$M_X[\xi(t)]$\} for
matrix processes, $\hat X_i(t) \equiv \omega_i'(t)\hat\varepsilon^i$,
$i=x,y,z$, where $\xi(t)=(\xi_x(t),\xi_y(t), \xi_z(t))$ is the vector
of trial functions \footnote{The moment generating functional is an
alternative means of specifying the probability distributions of
stochastic processes.  The coefficients in the expansion of
$M_X[\xi(t)]$ in powers of the trial function $\xi(t)$ define the
moments of the process.}.  Then, the average can be ``absorbed''
inside the exponential function by expressing $M_X[\xi(t)]$ through a
{\it cumulant generating functional} \{$K_X[\xi(t)]$\} in terms of
{\it cumulant averages} denoted here by a double bar (see
Ref.~\cite{Kubo:Cumulants} for a discussion of the cumulant expansion
methods used here):
\begin{eqnarray}
&& \ev{\hat U'(t,0)} =
	\ev{\mathcal T \exp\left\{
	\sum_i \hat\varepsilon^i\int_0^t\omega_i '(\tau)d\tau
	\right\}} =
	M_{X}[\xi(t)] \Bigg|_{
	\scriptsize\begin{aligned}\xi(t) &\equiv 1 \\[-4pt]
	\hat X_i &\equiv \omega_i'\hat\varepsilon^i\end{aligned}
	} \equiv  \nonumber\\
&&\equiv\mathcal T\exp\left\{ K_{\omega\hat\varepsilon}
    [ 1] \right\} = \mathcal T\exp\left\{ \ev{\ev{
	\mathcal T\exp\left(
	\sum_i \hat\varepsilon^i\int_0^t\omega_i '(\tau)d\tau
	\right)-\hat 1}}
	\right\} =\nonumber\\
&&=\mathcal T \exp\left\{
	\sum_{i,j}\hat\varepsilon^i\hat\varepsilon^j
	\int_0^t d\tau_1\int_0^{\tau_1}d\tau_2\, \ev{\ev{
	\omega_i'(\tau_1)\omega_j'(\tau_2)
	}}\right\} .
\end{eqnarray}
In the last step, we took advantage of the fact that all cumulants
beyond the second vanish for Gaussian processes, and
$\ev{\ev{\omega_i'(t)\omega_j'(t')}}=\ev{\omega_i'(t)\omega_j'(t')}$
if the average is zero.

Substituting the correlation functions given by
Eqs.~(\ref{eq:kappa_XY}), (\ref{eq:kappa_XX}) and (\ref{eq:kappa_ZZ}),
we find the following expession for the cumulant generating functional
$K$ evaluated at $\xi(t) \equiv 1$,
\begin{equation}\label{eq:K}
K_{\omega'\hat\varepsilon}[1]=(\hat\varepsilon^z)^2 \int_0^t d\tau \,
\gamma(\tau) +
\left( (\hat\varepsilon^x)^2 + (\hat\varepsilon^y)^2\right)
\int_0^t d\tau \, \gamma_+(\tau) +
\left(\hat\varepsilon^x\hat\varepsilon^y - \hat\varepsilon^y
\hat\varepsilon^x \right) \int_0^t d\tau \, \gamma_-(\tau) ,
\end{equation}
where the rates $\gamma$ are given by integrals of the correlation
functions:
\begin{eqnarray}
\gamma(t) &=& \int_0^t d\tau \, \ev{\omega_z'(\tau)\omega_z'(0)} =
\int_0^t d\tau \; \kappa(\tau),\label{eq:gamma}\\
\gamma_+(t) &=& \int_0^t d\tau \, \ev{\omega_x'(\tau)\omega_x'(0)}
= \int_0^t d\tau \, \ev{\omega_y'(\tau)\omega_y'(0)} =
 \int_0^t d\tau \; \kappa(\tau)\cos(\Omega_0 \tau ),\label{eq:gamma+}\\
\gamma_-(t) &=& \int_0^t d\tau \, \ev{\omega_x'(\tau)\omega_y'(0)}
= -\int_0^t d\tau \, \ev{\omega_y'(\tau)\omega_x'(0)} =
 \int_0^t d\tau \; \kappa(\tau)\sin(\Omega_0 \tau) .\label{eq:gamma-}
\end{eqnarray}
$K$ can be further simplified if we take into account properties of
the rotation generators $\hat\varepsilon^i$.  From the commutation
relations (\ref{eq:EpsCommRel}), the commutator in the third term of
Eq.  (\ref{eq:K}) is equal to $\hat\varepsilon^z$.  It is easy to
verify that the sum of squares of $\hat\varepsilon^x$ and
$\hat\varepsilon^y$ in the second term of $K$ can be written as
$\hat\varepsilon^{x\,2}+\hat\varepsilon^{y\,2}=-(2 \,\hat 1 +
\hat\varepsilon^2)$, hence, the parts of Eq.~(\ref{eq:K}) commute with
each other.  Thus, the average evolution matrix, which is equal to the
exponential of $K$, factorizes into three parts,
\begin{align}
\ev{\hat U'(t,0)}
& = \mathcal T\exp\{ K_{\omega'\hat\varepsilon}[1]\} =\exp\left\{
\hat\varepsilon^{z\,2}\int_0^td\tau\,\gamma(\tau)
-\left(2\hat 1+\hat\varepsilon^{z\,2}\right)\int_0^t d\tau\,
\gamma_+(\tau) + \hat\varepsilon^z\int_0^td\tau\,\gamma_-(\tau)
\right\} =\nonumber\\
&=\underbrace{
\left(\begin{array}{ccc}
e^{-\Gamma(t)}&0&0\\0&e^{-\Gamma(t)}&0\\0&0&1\\
\end{array}\right)
}_{\substack{\text{decay caused by}\\\text{the fluctuations in $z$}}}
\underbrace{
\left(\begin{array}{ccc}
e^{-\Gamma_+(t)}&0&0\\0&e^{-\Gamma_+(t)}&0\\0&0&e^{-2\Gamma_+(t)}\\
\end{array}\right)
}_{\substack{\text{decay caused by} \\\text{the fluctuations in $xy$ plane}}}
\underbrace{
\left(\begin{array}{ccc}
\cos\Gamma_-(t)&-\sin\Gamma_-(t)&0\\
\sin\Gamma_-(t)&\cos\Gamma_-(t)&0\\
0&0&1\\
\end{array}\right)
}_{\substack{\text{precession frequency modification}
\\\text{due to fluctuations in $xy$ plane}}}
,\label{eq:Avg_U}
\end{align}
where
\begin{eqnarray}
\Gamma_+(t) & = &
\int_0^t d\tau\,\gamma_+(\tau)=
\left(\frac{\omega_0^2 \tau_c^2}{1+\Omega_0^2\tau_c^2}\right)
\left[
 \frac{t}{\tau_c} +
 \left(\frac{1-\Omega_0^2\tau_c^2}{1+\Omega_0^2\tau_c^2}\right)\left(
 e^{-\frac{t}{\tau_c}}\cos\Omega_0 t - 1\right)
 -\left(\frac{2\Omega_0\tau_c}{1+\Omega_0^2\tau_c^2}\right)
 e^{-\frac{t}{\tau_c}} \sin\Omega_0 t
\right], \label{eq:Gamma+} \\
\Gamma_-(t) &=& \int_0^t d\tau \, \gamma_-(\tau)=
\left(\frac{\omega_0^2 \tau_c^2}{1+\Omega_0^2\tau_c^2}\right)
\left[
 \Omega_0 t +
 \left(\frac{1-\Omega_0^2\tau_c^2}{1+\Omega_0^2\tau_c^2}\right)
   e^{-\frac{t}{\tau_c}}
  \sin\Omega_0 t
 +\left(\frac{2\Omega_0\tau_c}{1+\Omega_0^2\tau_c^2}\right)
   \left(e^{-\frac{t}{\tau_c}}\cos\Omega_0 t - 1\right)
\right] ,  \label{eq:Gamma-} \\
\Gamma(t) &=& \int_0^td\tau\,\gamma(\tau)
 = \omega_0^2 \tau_c \left[ t - \tau_c \left(
 1-e^{-t/\tau_c} \right) \right] .  \label{eq:Gamma}
\end{eqnarray}
The first term on the right hand side of Eq.~(\ref{eq:Avg_U})
describes the decay caused by the fluctuations in the direction
parallel to the applied field $\vec\Omega_0=\Omega_0 \hat e_z$.  These
fluctuations affect only spin components perpendicular to the $z$
axis, and the decay rate $\Gamma(t)$ does not depend on the intensity
of the constant field.  The second term originates from the
fluctuations in the $x y$ plane, which is perpendicular to the applied
field.  Fluctuations in the plane couple to all the components of the
spin and cause their decay.  The key feature is that the rate of decay
due to these fluctuations is dependent on $\Omega_0$.  Finally, the
third term, also related to the fluctuations in the $x y$ plane,
describes the modification of the precession frequency of the system
caused by the noise.

\begin{figure}[]
\centering
\includegraphics[
angle=0,width=0.5\textwidth]{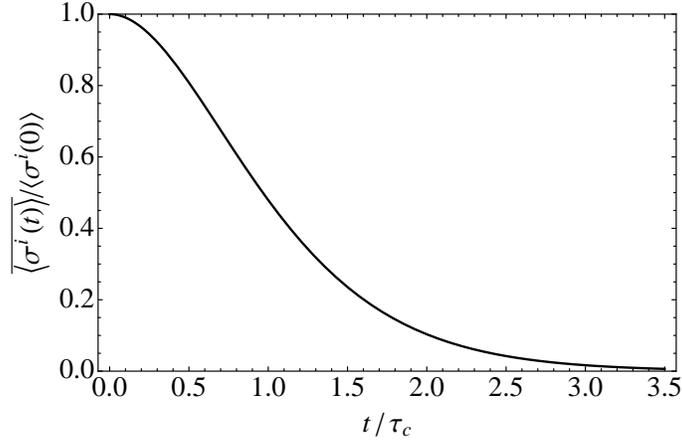}
\caption{The expectation value of the spin, $\langle\ev{{\vec S}
(t)}\rangle$, subject to Gaussian colored noise, versus time for
$\Omega_0 = 0$ and $\omega_0 \tau_c = 1$.  The analytic formula for
$\langle\ev{{\vec S} (t)}\rangle$ is given in
Eq.~(\ref{Eq:sigma_iso}).}
\label{Fig_Gaussian_CN_no_B_0}
\end{figure}

The averaged expectation values of the spin are given by:
\begin{eqnarray}
\langle\ev{ S_x (t)}\rangle &=& e^{-\left[\Gamma(t) +
\Gamma_+(t)\right]}
\left\{
 \cos\Big[\Omega_0 t + \Gamma_-(t)\right]\langle S_x\rangle
 -\sin\left[\Omega_0 t + \Gamma_-(t)\right]\langle S_y\rangle
\Big\} , \label{Eq:sigma_x} \\
\langle\ev{ S_y (t) }\rangle &=& e^{-\left[\Gamma(t)+\Gamma_+(t)\right]}
\Big\{
 \sin\left[\Omega_0 t + \Gamma_-(t)\right]\langle  S_x\rangle
 +\cos\left[\Omega_0 t + \Gamma_-(t)\right]\langle S_y\rangle
\Big\}, \label{Eq:sigma_y} \\
\langle \ev{ S_z (t)} \rangle &=& e^{-2\Gamma_+(t)}\langle S_z\rangle
\label{Eq:sigma_z} .
\end{eqnarray}
In the limit $\Omega_0 \to 0$, we obtain $\Gamma_+(t) \to \Gamma(t)$ and
$\Gamma_-(t) \to 0$, hence Eqs.~(\ref{Eq:sigma_x})-(\ref{Eq:sigma_z})
reduce to
\begin{equation} \label{Eq:sigma_iso}
    \langle\ev{{\vec S} (t)}\rangle \xrightarrow[\Omega_0\to 0]{}
    e^{-2\Gamma(t)} \langle {\vec S} \rangle .
\end{equation}
Substituting Eq.~(\ref{eq:Gamma}) into (\ref{Eq:sigma_iso}) yields the
analytic form (\ref{Eq:sigma_iso}) of the isotropic decay of the
average spin, regardless of the initial state of the spin, as can be
seen in Fig.~\ref{Fig_Gaussian_CN_no_B_0}.  For times $t \gg \tau_c$,
the decay is simply exponential with the rate of decay given by
$\Gamma(t) \approx \omega_0^2 \tau_c t$.  For times $t \ll \tau_c$,
the rate of decay is $\Gamma(t)\approx \omega_0^2 t^2/2$.  In the
absence of an external field, the fluctuations in all three directions
remain independent and the average evolution matrix decomposes in the
following way:
\begin{equation}
\ev{\hat U(t,0)}\Big|_{\Omega_0=0}=
\underbrace{
\left(\begin{array}{ccc}
e^{-\Gamma(t)}&0&0\\
0&e^{-\Gamma(t)}&0\\
0&0&1\\
\end{array}\right)
}_{\text{fluctuations in $z$}}
\underbrace{
\left(\begin{array}{ccc}
e^{-\Gamma(t)}&0&0\\
0&1&0\\
0&0&e^{-\Gamma(t)}\\
\end{array}\right)
}_{\text{fluctuations in $y$}}
\underbrace{
\left(\begin{array}{ccc}
1&0&0\\
0&e^{-\Gamma(t)}&0\\
0&0&e^{-\Gamma(t)}\\
\end{array}\right)
}_{\text{fluctuations in $x$}}.
\end{equation}
Here the first part comes from fluctuations in $z$ and causes the
decay of the $x$ and $y$ components of the spin, the second part comes
from fluctuations in $y$ and causes the decay of the perpendicular
components, and the third part originates from fluctuations in $x$
direction causes the decay of the $y$ and $z$ components of the spin.

\begin{figure}[]
\centering
\includegraphics[
angle=0,width=0.5\textwidth]{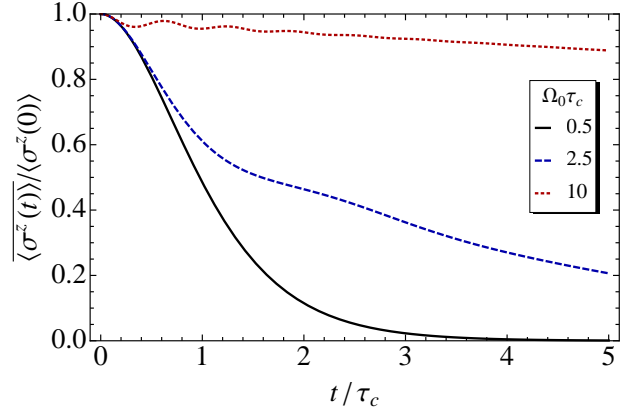}
\caption{$\langle\ev{  S_z (t)} \rangle$ subject to Gaussian
colored noise versus time for nonvanishing constant $\Omega_0$ (see
legend) and $\omega_0 \tau_c = 1$.  Equation (\ref{Eq:sigma_z}) gives
the analytic formula for $\langle \ev{ S_z(t)} \rangle$.}
\label{Fig_Gaussian_CN_sigma_z_B_0}
\end{figure}

Now let us consider finite $\Omega_0$.  The dependence of $\langle
\ev{S_z (t)} \rangle$, the component parallel to the external field
$\vec\Omega_0$, versus time for three values of dimensional parameter
$\Omega_0 \tau_c$ is plotted in
Fig.~\ref{Fig_Gaussian_CN_sigma_z_B_0}.  As we discussed previously,
$z$ component of the average spin decays due to fluctuations in the
$xy$ plane, which {\it are} affected by the presence of the constant
field.  For $\Omega_0 \tau_c \ll 1$, the quantity $\langle\ev{ S_z (t)}
\rangle$ is clearly very similar to the results plotted in
Fig.~\ref{Fig_Gaussian_CN_no_B_0}, since $\Gamma_+(t) \to \Gamma(t)$.
As $\Omega_0 \tau_c$ increases, the decay of $\langle\ev{ S_z (t)}
\rangle$ is significantly slowed, and $\Gamma_+(t) \to 0$ in the
limit of $\Omega_0 \tau_c \to \infty$, i.e., the decay is fully
suppressed.

\begin{figure}[]
\begin{minipage}[b]{0.45\linewidth}
\centering
\includegraphics[
angle=0,width=\textwidth]{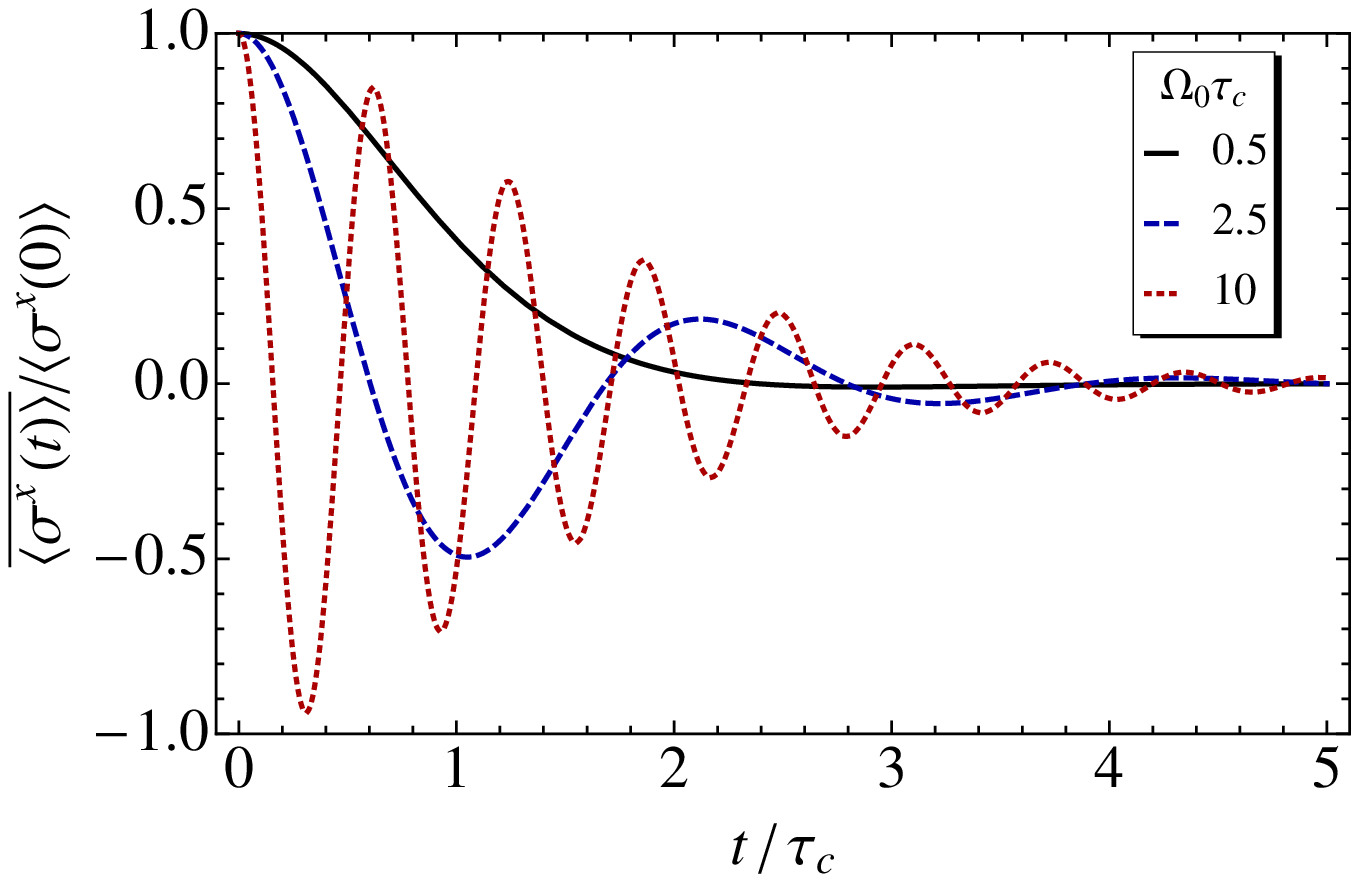}
\caption{The average value $\langle  S_x \rangle$, when the spin
is subject to Gaussian colored noise for nonvanishing constant
$\Omega_0$ (see legend) and $\omega_0 \tau_c = 1$.  Equation
(\ref{Eq:sigma_x}) together with initial condition $\langle
 S_y(0) \rangle = 0$ yield the analytic formula for the function
plotted.  The average value $\langle  S_y \rangle$ for initial
condition $\langle  S_x(0) \rangle = 0$ is identical to the
results shown here.}
\label{Fig_Gaussian_CN_sigma_x_B_0}
\end{minipage}%
\hspace{.5cm}\begin{minipage}[b]{0.45\linewidth}
\centering
\includegraphics[
angle=0,width=\textwidth]{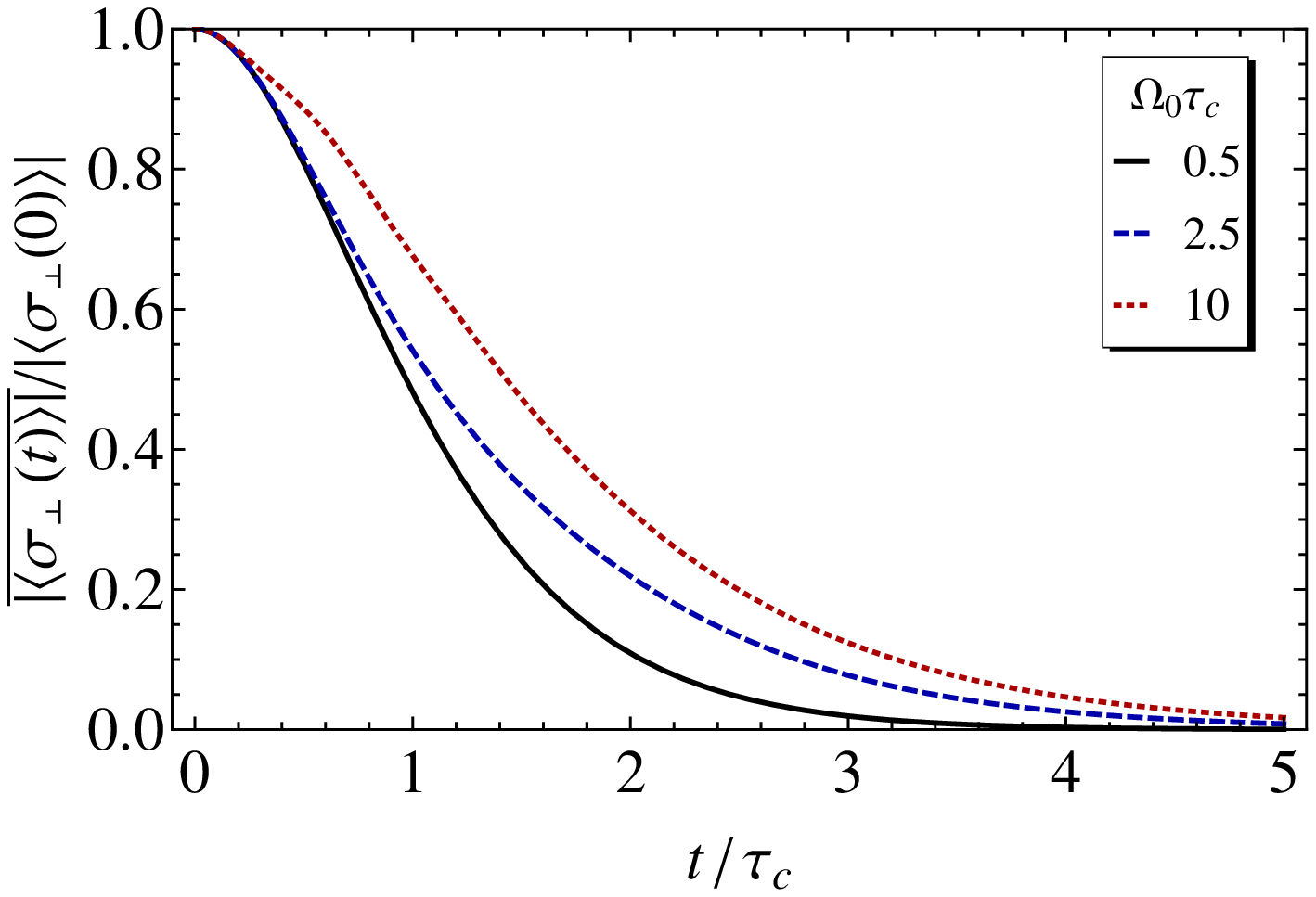}
\caption{Average value of $\left|\langle\ev{ S_\perp(t)
}\rangle\right| = \sqrt{\langle\ev{ S_x(t)}\rangle^2 +
\langle\ev{ S_y(t)}\rangle^2}$ subject to Gaussian colored noise
for nonvanishing constant $\Omega_0$ (see legend) and $\omega_0 \tau_c
= 1$.  Equations (\ref{Eq:sigma_x}) and (\ref{Eq:sigma_y}) yield the
analytic formula for the function plotted.}
\label{Fig_Gaussian_CN_sigma_perp_B_0}
\end{minipage}
\end{figure}

Figure~\ref{Fig_Gaussian_CN_sigma_x_B_0} plots $\langle\ev{ S_x (t)}
\rangle$ versus time for the case when $\langle S_y(0) \rangle = 0$
and Fig.~\ref{Fig_Gaussian_CN_sigma_perp_B_0} presents the time
dependence of the length of the vector $\langle\ev{{\vec
S}_\perp(t)}\rangle = (\langle\ev{ S_x(t)}\rangle, \langle
\ev{S_y(t)}\rangle, 0)$.  The decay of the components of the average spin
perpendicular to the constant field $\vec\Omega_0$ is not only induced
by the fluctuations in the $xy$ plane, but also by those in the $z$
direction [see Eq.~(\ref{eq:Avg_U})].  We have previously seen that
the effects of the fluctuations in $x$ and $y$ are inhibited if
$\Omega_0\tau_c$ is large enough.  However, the influence of the
fluctuations in the direction parallel to applied field is unperturbed
by the presence of $\Omega_0$.  As long as the $z$ component of the
field is fluctuating the perpendicular component of the average spin
will decay to zero for any value of $\Omega_0 \tau_c$ as $t/\tau_c \to
\infty$.  For small $\Omega_0 \tau_c$ the decay is, again, similar to
the isotropic case shown in Fig.~(\ref{Fig_Gaussian_CN_sigma_z_B_0}),
but when $\Omega_0 \tau_c$ increases, the decay is modulated by
oscillations which result from the precession of the spin about the
external field ${\vec \Omega}_0 $ with frequency modified by the
fluctuations.  For $t \gg \tau_c$, the oscillation frequency is given
by $\Omega_0 \left(1 + \frac{\omega_0^2 \tau_c^2} {1 +
\Omega_0^2\tau_c^2} \right)$ [see Eq.~(\ref{eq:Gamma-})].  When
$\Omega_0 \tau_c \gg 1$ the frequency of precession tends to
$\Omega_0$ as the frequency modification due to fluctuations,
$\Gamma_-(t)$ vanishes, in a similar manner as that of $\Gamma_+(t)$.

For spin 1/2 system (a TLS), the decay of $\langle\ev{ S_x (t)}
\rangle$ and $\langle\ev{ S_y (t) }\rangle$ correspond to the decay
of the coherence of the system, whereas the decay of $\langle\ev{ S_z
(t) }\rangle$ corresponds to the vanishing of population imbalance.
If the constant field is taken along $x$ (as opposed to along $z$),
decay of $\langle \ev{S_x (t)} \rangle$ is suppressed, and the
decoherence time $T_2 \sim \tau_c \Omega_0^2/\omega_0^2$, which can be
made arbitrarily large (suppressed decay) by increasing $\Omega_0$.

The averaged spin vector solution obtained in this section can be
extended to the case of a time dependent deterministic field.  If
$\Omega_0$ is a function of time then the frequency of oscillations in
the $\cos$ and $\sin$ functions appearing in the definitions of the
decay rates (\ref{eq:gamma+}) and (\ref{eq:gamma-}) are to be replaced
by an integral, $\Omega_0 t \to \int_0^t \Omega_0(t')dt'$.

\section{The White Noise Limit} \label{Sec:white-noise}

The white noise limit of colored noise involves going to the regime of
extremely rapid fluctuations of the field, i.e., vanishing correlation
time $\tau_c$.  Unfortunately, simply taking $\tau_c \to 0$ does not
yield the correct limit.  If we put $\tau_c = 0 $ in
Eq.~(\ref{eq:Gamma}), we obtain that the decay rate of the average
spin equals zero, and fluctuations do not affect the evolution of the
spin.  The proper limit can be obtained if the strength of
fluctuations, as given by the variance of the field, $\ev{\omega(t)^2}
= \omega_0^2$, is increased as the correlation time $\tau_c$ goes to
zero, so that $\omega_0^2 \tau_c$ remains constant.  Thus, we arrive
at the limit when the correlation function of the colored noise
becomes proportional to a Dirac delta function,
\begin{equation}
    \ev{\omega_i(t)\omega_j(t')} = \omega_0^2 \, e^{-\frac{|t-t'|}
    {\tau_c}} \delta_{ij} = 2\omega_0^2 \tau_c \,
    \left(\frac{1}{2\tau_c} \, e^{-\frac{|t-t'|}{\tau_c}} \right)
    \delta_{ij} \xrightarrow[\tau_c \to 0]{\omega_0^2\tau_c \to D} 2 D
    \delta(t-t')\delta_{ij} .
\end{equation}
In the limit of $\kappa(\tau)\to 2 D \delta(\tau)$ the decay rates
$\Gamma_{+}(t) = \Gamma(t) = 2Dt$ and $\Gamma_{-}(t) = 0$.  The decay
of the expectation value of the spins are then isotropic and are not
affected by the presence of a constant field $\vec \Omega_0$, i.e.,
\begin{equation}  \label{Eq:white_noise_expect}
    \langle{\ev{\vec S (t)}}\rangle = \left(\begin{array}{ccc}
    e^{-\Gamma(t)-\Gamma_+(t)}\cos\Gamma_-(t) &
    -e^{-\Gamma(t)-\Gamma_+(t)}\sin\Gamma_-(t) & 0 \\
    e^{-\Gamma(t)-\Gamma_+(t)}\sin\Gamma_-(t) &
    e^{-\Gamma(t)-\Gamma_+(t)}\cos\Gamma_-(t) & 0\\
    0 & 0 & e^{-2\Gamma_+(t) }\\ \end{array}\right)
    \, \hat R(\Omega_0 t,z) \, \langle{\vec S}\rangle \to e^{-4 D
    t}\hat R(\Omega_0 t,z)\,\langle{\vec S}\rangle .
\end{equation}

Thus, the spin decay is isotropic, purely exponential with decay rate
$4 D$, and is independent of $\Omega_0$.  Moreover, the precession due
to $\vec\Omega_0$ is unaffected by the fluctuations.  In the white
noise limit, the precession due to deterministic part of the field and
the decoherence due to field fluctuations are completely decoupled.

\section{Memory effects}  \label{Sec:Memory}

In the expression for $\langle{\ev{\vec S(t)}}\rangle$ for the case of
Gaussian colored noise, the decay rates of the spin depend on the
history of the fluctuating field through the integrals of correlation
functions.  But the decay rate for the white noise case does not
depend on history.  Based on this observation, we say that spin system
has {\it memory} if the fluctuating field driving it has finite
correlation time.  This definition is consistent with a rigorous
notion of memory, since ${\vec S}(t)$, understood as a stochastic
process, is Markovian~\cite{vanKampenBook} (i.e. memoryless) if and
only if the noise driving it is white noise.  Memory in the system has
important consequences, because it allows control of spin decay,
through the application of an external field.

The role of memory can be understood as follows.  Consider first the
case of $\Omega_0 = 0$.  Each realization of the fluctuating field
${\vec \omega}(t)$ determines corresponding realization of $\langle
\vec S(t)\rangle = \hat U(t,0)\langle\vec S\rangle$ via the equation
of motion, where $\vec\omega(t)$ defines an instantaneous rotation
axis for the spin vector.  Hence, each realization draws a trajectory
of the spin on a sphere of radius $|\langle \vec S \rangle|$.  The
evolution can be viewed in a simplified ``stroboscopic'' picture in
which, within a correlation time $\tau_c$, the axis $\vec\omega(t)$
does not change much, and subsequently, it jumps to a new random value
within a range on the order of $\omega_0$.  Eventually such a sequence
of rotations around randomly selected axes ``smears'' the trajectories
of the spin over the whole sphere.  This results in the decay of the
spin averaged over all realizations.

The situation is different when a constant field along, say, the
$z$-axis is present.  This becomes clear when we view the evolution in
the reference frame rotating with frequency $\Omega_0$ around the
$z$-axis.  Consider a particular realization of the fluctuating field
${\vec \omega}(t)$, and denote it in the rotating frame as ${\vec
\omega}'(t)$.  Since $\omega'_z(t) = \omega_z(t)$, the fluctuations in
the direction parallel to the constant field are unchanged when
shifted to the rotating frame.  On the other hand, the projection of
the fluctuating field on the plane perpendicular to $z$-axis,
$\vec\omega_\perp(t) = (\omega_x(t), \omega_y(t), 0)$, is modified
when viewed in the rotating frame.  Within a time of order the
correlation time $\tau_c$, the projection $\vec\omega_\perp'(t)$
revolves around the $z$-axis with frequency $\Omega_0$, and if
$\Omega_0 \tau_c \gg 1$, $\vec\omega_\perp'(t)$ is able to perform
many revolutions between jumps.  The effect of the jumps becomes
negligible in comparison with the systematic rotation.  Thus the
stochastic character of $\vec\omega'_\perp(t)$ is lost, and the only
contribution to the decay caused by fluctuations comes from
$\omega_z(t)$.  A similar effect provides stable motion of a spinning
top, and for that reason we will refer to it as a {\it gyroscopic
effect}.  If $\Omega_0 \tau_c \ll 1$, then revolution in between jumps
is negligible, so $\vec\omega'_\perp(t) \approx \vec\omega_\perp(t)$,
and by necessity the decay induced by the fluctuations must be almost
the same as if there was no constant field at all.

The suppression mechanism of the decay described above is inoperative
in the white noise limit where the correlation time vanishes, so the
fluctuations of the rotation axis are instantaneous.  Hence, the decay
caused by white noise cannot be affected by a constant field of finite
intensity.  Moreover, even if the strength of the fluctuations is
small in comparison to $\Omega_0$, the effect of fluctuations cannot
be neglected if the correlation time is too short.

Note that the important dimensionless parameter here is $\Omega_0
\tau_c$, which can be interpreted as an average angle of precession
due to the deterministic field within the time period of $\tau_c$.
Bearing this in mind, it is clear that the conclusion of this
paragraph remains valid even if the field $\Omega_0$ varies in time as
long as $\tau_c \langle \Omega_0(t)\rangle_{\tau_c} \equiv
\tau_c\left(\frac 1{\tau_c} \int_t^{t+\tau_c}\Omega_0(t')dt'\right)
\gg 1$.  Moreover, the reasoning used above shows that the same decay
suppression effects apply even for non-Gaussian noise as long as it
has non-vanishing correlation time.  We verify this assessment in the
following section.

\section{Non-Gaussian process: telegraph noise}\label{Sec:Telegraph}

In this section we present an example of a non-Gaussian process,
telegraph noise, i.e., dichotomic noise corresponding to a memoryless
continuous-time stochastic process that jumps between two distinct
values, sometimes called burst noise \cite{vanKampenBook,
GardinerBook}.  We consider telegraph noise in one component of the
field, $\omega_z(t)$.  The Rabi frequency of the fluctuating field,
$\omega_z(t)$, jumps randomly between $+ \omega_0$ and $- \omega_0$,
with a jump rate $w = (2 \tau_c)^{-1}$.  The average value and the
correlation function of such a process is given by
\begin{equation}
    \overline{\omega_z(t)} = 0 , \quad \overline{\omega_z(t)\omega_z(t')}
    = \omega_0^2 e^{-2w|t-t'|}=\omega_0^2 \, e^{-\frac{|t-t'|}{\tau_c}} .
\end{equation}
Because telegraph noise is non-Gaussian, the correlation function
$\kappa(t-t') =\omega_0^2 e^{-\frac{|t-t'|}{\tau_c}}$ does not contain all the
information regarding the process.

Telegraph noise can model the effect of a spin 1/2 impurity atom near
the spin \cite{Itakura_03, Cheng_08, Aharony_10}.  The impurity spin
has probabilities $p_+$ and $p_-$ for being in the spin-up and
spin-down state, and a hopping rate for transferring from one spin
state to the other.  Neglecting the back-action of the magnetic moment
on the impurity, this model reduces to telegraph noise.

\begin{figure}[h]
\centering
\includegraphics[ angle=0,width=0.5\textwidth]{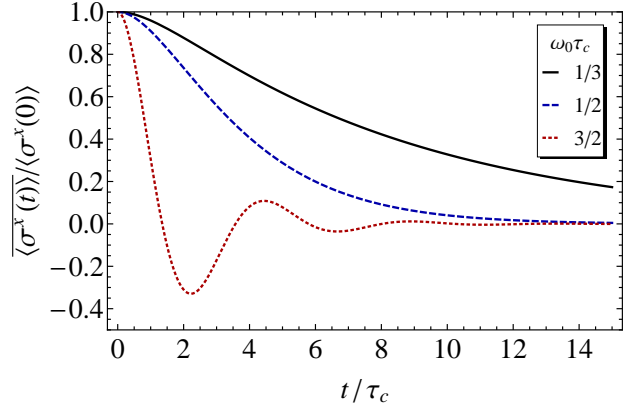}
\caption{$\langle \ev{ S_x (t)} \rangle$ versus time for a spin
subject to telegraph noise in only the $z$-component of the field,
$\omega_z(t)$, and $\Omega_0 = 0$.  $\langle \ev{S_y (t)} \rangle$ is
identical to $\langle \ev{ S_x (t)} \rangle$.  Nonvanishing values of
$\Omega_0$ simply cause precession of the spin about the $z$ axis.}
\label{Fig_Telegraph}
\end{figure}

Our goal is to find the average evolution matrix $\ev{\hat U(t,0)}$
for the telegraph noise case.  The technique we have used previously
to find the average does not work in the case of telegraph noise,
because unlike with Gaussian processes, the closed form of the
cumulant generating functional is not known.  Fortunately, it is still
possible to find an exact, analytical formula for $\ev{\hat U(t,0)}$
if the fluctuations are parallel to the constant field.  One way of
solving for $\ev{\hat U(t,0)}$ is to follow
Ref.~\cite{VanKampen_Tele}.  We can write the average evolution
operator as $\ev{\hat U(t,0)} = \ev{\hat U(t,0)}_+ + \ev{\hat
U(t,0)}_-$, where $\ev{\hat U(t,0)}_\pm$ are the averages conditional
on $\omega_z(t)$ being equal to $\pm\omega_0$.  These averages vary in
time because of equation of motion (\ref{eq:U}), and because
$\omega_z(t)$ jumps.  Thus,
\begin{eqnarray}
    \frac{d}{dt}\ev{\hat U(t,0)}_+ &=& (\Omega_0
    +\omega_0)\hat\varepsilon^z\ev{\hat
    U(t,0)}_+ - \frac{1}{2\tau_c}\ev{\hat
    U(t,0)}_+ + \frac{1}{2\tau_c}\ev{\hat U(t,0)}_-,\\
    \frac{d}{dt}\ev{\hat U(t,0)}_- &=& (\Omega_0
    -\omega_0)\hat\varepsilon^z\ev{\hat
    U(t,0)}_- - \frac{1}{2\tau_c}\ev{\hat
    U(t,0)}_- + \frac{1}{2\tau_c}\ev{\hat U(t,0)}_+ .
\end{eqnarray}
Adding and subtracting these equations, we obtain
\begin{eqnarray}
    \frac{d}{d t}\ev{\hat U(t,0)} &=& \Omega_0 \hat\varepsilon^z
    \ev{\hat U(t,0)} +\omega_0\hat\varepsilon^z\ev{\Delta \hat U(t,0)},
    \label{Eq:Av_Ev_Op} \\
    \frac{d}{d t}\ev{\Delta \hat U(t,0)} &= &
    \Omega_0\hat\varepsilon^z\ev{\Delta\hat U(t,0)}
    +\omega_0\hat\varepsilon^z\ev{\hat U(t,0)}-\frac{1}{\tau_c}
    \ev{\Delta\hat U(t,0)}, \label{Eq:D_Ev_Op}
\end{eqnarray}
where $\ev{\Delta \hat U(t,0)} = \ev{\hat U(t,0)}_+ - \ev{\hat
U(t,0)}_-$.  The next step is to eliminate $\ev{\Delta \hat U(t,0)}$
from Eq.~(\ref{Eq:Av_Ev_Op}) by substituting the solution to
Eq.~(\ref{Eq:D_Ev_Op}), with the initial condition $\ev{\hat
U(0,0)}_\pm = \hat 1/2$.  This yields a closed equation for average
evolution matrix:
\begin{equation}\label{eq:Tele_Eq}
   \frac{d}{dt}\ev{\hat U(t,0)} = \Omega_0\hat\varepsilon^z \ev{\hat
   U(t,0)} + \omega_0^2\int_0^t dt' \, e^{-t'/\tau_c}
   (\hat\varepsilon^z)^2 \ev{\hat U(t-t',0)} .
\end{equation}
Note the particular form of the ``memory integral'' in the equation
(\ref{eq:Tele_Eq}), where the dynamical variable is evaluated at times
previous to $t$.  In this case the system clearly has memory and it
remembers not only the history of the noise but also its own history.
It is easy to check that the solution to this equation that satisfies
the initial condition $\ev{\hat U(0,0)} = \hat 1$ is given by
\begin{equation}\label{eq:TeleU}
    \ev{\hat U(t,0)} = e^{-t/2\tau_c}\left[
    \cosh\left(\hat\gamma\frac{t}{2\tau_c}\right) +
    \hat\gamma^{-1}\sinh\left(\hat\gamma\frac{t}{2\tau_c}\right)
    \right]\hat R(\Omega_0 t, z),
\end{equation}
where
\begin{equation}
    \hat\gamma = \left(\begin{array}{ccc}
    \sqrt{1-4\omega_0^2\tau_c^2} & 0 & 0 \\
    0 & \sqrt{1-4\omega_0^2\tau_c^2} & 0 \\
    0 & 0 & 1 \\
    \end{array}\right).
\end{equation}
The average expectation values of the spin are obtained by applying
the average evolution matrix (\ref{eq:TeleU}) to the initial vector of
spin operators, $\langle\ev{\vec S(t)}\rangle=\ev{\hat
U(t,0)}\langle{\vec S}(0)\rangle$, and this yields,
\begin{eqnarray}
    \langle \ev{ S_x (t)}\rangle &=& e^{-\frac{t}{2\tau_c}}
    \left[ \cosh\left(\sqrt{1-4\omega_0^2\tau_c^2}\frac{t}{2\tau_c}\right)
    + \frac{\sinh\left(\sqrt{1-4\omega_0^2\tau_c^2}\frac{t}{2\tau_c}\right)}
    {\sqrt{1-4\omega_0^2\tau_c^2}} \right]
    \left[  \cos(\Omega_0 t) \langle S_x\rangle
     -\sin(\Omega_0 t )\langle S_y\rangle \right] ,
     \label{eq:tele_sigma_x} \\
     \langle\ev{  S_y (t) }\rangle &=& e^{-\frac{t}{2\tau_c}}
     \left[\cosh\left(\sqrt{1-4\omega_0^2\tau_c^2}\frac{t}{2\tau_c}\right)
     +\frac{\sinh\left(\sqrt{1-4\omega_0^2\tau_c^2}\frac{t}{2\tau_c}\right)}
     {\sqrt{1-4\omega_0^2\tau_c^2}} \right]
     \left[\cos(\Omega_0 t) \langle S_y\rangle
     + \sin(\Omega_0 t )\langle S_x\rangle \right] ,
     \label{eq:tele_sigma_y} \\
     \langle\ev{  S_z (t) }\rangle &=& \langle S_z\rangle
     \label{eq:tele_sigma_z} .
\end{eqnarray}
Neither constant field $\Omega_0$, nor field fluctuations
$\omega_z(t)$ couple to the $z$ component of the spin, hence it
remains constant throughout the evolution.  The effects of $\Omega_0$
and $\omega_z(t)$ on the components of the spin in $xy$ plane are
simply superposed: constant field causes trivial precession around $z$
axis, while the noise makes the $x$ and $y$ spin components decay on
average.  Moreover, if the constant field and the fluctuating field
are parallel, the gyroscopic effect can not be observed.  The only
memory effect present in this case is a quadratic behavior in $t$ of
the decay rate at $t\ll \tau_c$.  Figure~\ref{Fig_Telegraph} plots
$\langle\ev{ S_x (t) }\rangle$ versus time for telegraph noise in
$\omega_z(t)$ for the case of no static external magnetic present.  As
$\omega_0 \tau_c$ increases, $\langle\ev{ S_x (t) }\rangle$ decays
more quickly, but for $\omega_0 \tau_c > 1/2$, the square root
$\sqrt{1-4\omega_0^2\tau_c^2}$ becomes imaginary and the solution
changes form from pure decay to decay with oscillations.

\begin{figure}[h]
\centering
\includegraphics[ angle=0,width=0.5\textwidth]{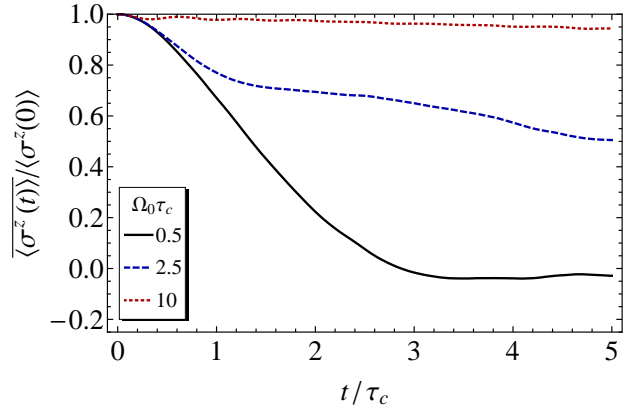}
\caption{$\langle \ev{ S_z (t) }\rangle$ versus time for a spin
subject to telegraph noise only in the $x$ direction, with
$\omega_0\tau_c = 1$.  The deterministic magnetic field with Rabi
frequency $\Omega_0$ is taken to be in the $z$ direction, so it is
perpendicular to the noise.  As the parameter $\Omega_0\tau_c$ is
increased (see legend), suppression of the decay of the spin
increases.}
\label{Fig_Telegraph_Perp}
\end{figure}

It is of interest to see what happens when noise is present in the
component of the field perpendicular to the constant field $\Omega_0$,
e.g., in $\omega_x(t)$.  Unfortunately, in this case, it is not
possible to obtain a closed form expression for the average evolution
matrix.  According to the previous discussion, one would expect that,
since telegraph noise is colored noise (the correlation time $\tau_c$
is finite), the gyroscopic effect should work in a fashion similar to
the Gaussian colored noise example.  This is indeed the case.
Numerical solutions of $\langle\ev{ S_z (t)} \rangle$ versus time are
plotted in Fig.~\ref{Fig_Telegraph_Perp} and suppression of the decay
is more efficient with increasing $\Omega_0\tau_c$.

\section{Summary and Conclusion}   \label{Sec:Summary}

We introduced a stochastic model for spin in the presence of a
deterministic magnetic field and a fluctuating magnetic field (noise),
and we derived stochastic equations of motion for the spin vector
$\langle{\vec S}(t)\rangle$.  The environmental degrees of freedom,
represented by the fluctuations of the field, are eliminated by taking
the average over the fluctuations to obtain $\langle \ev{{\vec S}(t)}
\rangle$.  We made the {\it external noise} assumption, wherein no
back-action of the system on the environment is present.  The most
pronounced consequence of this assumption is that the system does not
go into equilibrium with a thermal environment, but instead goes to
the most democratic density matrix, i.e., the system tends towards the
completely mixed state, and $\langle \ev{{\vec S}(t)} \rangle \to 0$
as $t \to \infty$.  This is in contradistinction to the case of noise
experienced by a system in contact with a bath wherein the bath
affects the system {\em and} vice versa.  The lack of mutual
interaction (back-action) means that the fluctuation-dissipation
theorem \cite{Callen_Welton_51} cannot be applied, and a thermal
equilibrium state of the system is not obtained at large times.  If
back-action on the bath due to the system is present, the source of
the fluctuations and the dissipation is the same, they are connected
through fluctuation-dissipation theorem, and thermal equilibrium of
the system must result.

We explicitly considered Gaussian colored noise, Gaussian white noise
and non-Guassian telegraph noise.  From our studies we conclude that
if the system has memory, i.e., it has a finite correlation time, then
the system can be manipulated by means of an external magnetic field,
and the decoherence induced by the fluctuations can be significantly
suppressed.  However, in the white noise limit, when the correlation
function tends to a Dirac delta function, the decoherence cannot be
controlled in this way.  We introduced simple and intuitive
considerations for why the suppression of decay can be achieved for
any stochastic process as long as the correlation time is long enough.


\begin{acknowledgments}
This work was supported in part by grants from the Israel Science
Foundation (No.~2011295), and the James Franck German-Israel
Binational Program.  P.S. acknowledges the Foundation for Polish
Science International PhD Projects Programme co-financed by the EU
European Regional Development Fund, and the National Science Centre
DEC-2011/03/D/ST2/00200.  M.T. acknowledges financial support of the
National Science Centre.  We are grateful Doron Cohen and Marek
Ku\'{s} for useful conversations related to this work.
\end{acknowledgments}


\begin{thebibliography}{99}
\bibitem{No_indep_param}
The number of independent parameters needed to characterize
the density matrix of a spin $S$ system is $(2S+1)^2 -1$, so for $S =
1/2$ the three quantities $\langle S_i \rangle$ are sufficient, but
for higher $S$, the number of parameters exceeds the number of
expectation values of the spin components.
\end{thebibliography}
\end{document}